\documentclass[conference]{IEEEtran}
\IEEEoverridecommandlockouts
\usepackage{cite}
\usepackage{amsmath,amssymb,amsfonts}
\usepackage{algorithmic}
\usepackage{graphicx}
\usepackage{textcomp}
\usepackage{multirow}
\usepackage{array}
\usepackage{cite}
\usepackage{amsmath,amssymb,amsfonts}
\usepackage{algorithmic}
\usepackage{graphicx}
\usepackage{textcomp}
\usepackage{multirow}
\usepackage{array}
\usepackage{tikz}

\makeatletter
\def\set@curr@file#1{%
  \begingroup
    \escapechar\m@ne
    \xdef\@curr@file{\expandafter\string\csname #1\endcsname}%
  \endgroup
}
\def\quote@name#1{"\quote@@name#1\@gobble""}
\def\quote@@name#1"{#1\quote@@name}
\def\unquote@name#1{\quote@@name#1\@gobble"}
\makeatother

\def\BibTeX{{\rm B\kern-.05em{\sc i\kern-.025em b}\kern-.08em
    T\kern-.1667em\lower.7ex\hbox{E}\kern-.125emX}}
\begin{document}

\title{A Novel Approach to Classify Natural Grasp Actions by Estimating Muscle Activity Patterns from EEG Signals
\footnote{{\thanks{\hrule This research was partly supported by an Institute of Information \& Communications Technology Planning \& Evaluation (IITP) grant, funded by the Korean government (No. 2017-0-00432, Development of Non-Invasive Integrated BCI SW Platform to Control Home Appliances and External Devices by User’s Thought via AR/VR Interface) and partly funded by an Institute of Information \& Communications Technology Planning \& Evaluation (IITP) grant funded by the Korean government (No. 2017-0-00451, Development of BCI based Brain and Cognitive Computing Technology for Recognizing User’s Intentions using Deep Learning).}
}}
}

\author{\IEEEauthorblockN{Jeong-Hyun Cho$^1$, Ji-Hoon Jeong$^1$, Dong-Joo Kim$^1$, and Seong-Whan Lee$^{1,2}$}
\IEEEauthorblockA{$^1$Department of Brain and Cognitive Engineering, Korea University, Seoul, Republic of Korea \\
$^2$Department of Artificial Intelligence, Korea University, Seoul, Republic of Korea\\
jh$\_$cho@korea.ac.kr, jh$\_$jeong@korea.ac.kr, dongjookim@korea.ac.kr, sw.lee@korea.ac.kr}
}


\maketitle

\begin{abstract}
Developing electroencephalogram (EEG) based brain-computer interface (BCI) systems is challenging. In this study, we analyzed natural grasp actions from EEG. Ten healthy subjects participated in this experiment. They executed and imagined three sustained grasp actions. We proposed a novel approach which estimates muscle activity patterns from EEG signals to improve the overall classification accuracy. For implementing, we have recorded EEG and electromyogram (EMG) simultaneously. Using the similarity of the estimated pattern from EEG signals compare to the activity pattern from EMG signals showed higher classification accuracy than competitive methods. As a result, we obtained the average classification accuracy of 63.89$\pm$7.54\% for actual movement and 46.96$\pm$15.30\% for motor imagery. These are 21.59\% and 5.66\% higher than the result of the competitive model, respectively. This result is encouraging, and the proposed method could potentially be used in future applications, such as a BCI-driven robot control for handling various daily use objects.
\end{abstract}

\begin{small}\textbf{\textit{Keywords-brain-computer interface; electroencephalogram; electromyogram; robotic arm; motor imagery; hand grasping motion}\\}
\end{small}

\section{Introduction}
Decoding electroencephalogram (EEG) based brain-computer interfaces (BCIs) is a challenging task. Even with its difficulties, the BCIs are promising tools for detecting user intention and controlling robotic devices such as upper limb prosthesis \cite{pfurtscheller1997eeg, kim2016commanding, kim2014decoding}. Many research groups use EEG-based BCI because of its cost-effectiveness, convenience \cite{gilja2015clinical, kim2016commanding, kim2014detection, kim2014decoding}, and potentials \cite{kam2013non, kwak2017convolutional, jeong2019trajectory}. At the same time, improving the decoding accuracy of the BCI system is one of the major interests of many researchers \cite{suk2016deep, yeom2014efficient}.  
Among the many other motor-related EEG studies, we focused on hand movements. The hands are uniquely related to more dynamic brain activity than other extremities so we can acquire the EEG signals in the various amounts and aspects. About the decoding of movement in the hands and upper extremities, three related studies inspired our research. Schwarz et al. \cite{schwarz2017decoding} tried to decode natural reach and grasp actions from human EEG. They attempted to identify three different executed reach and grasp actions, namely lateral, pincer, and palmar grasp, utilizing EEG neural correlates.

Other research groups had slightly different approaches. Ofner et al. \cite{ofner2017upper} had encoded single upper limb movements in the time-domain of low-frequency EEG signals. The primary goal of the experiment was to classify six different actions, and those are elbow flexion, extension, hand grasp, spread, wrist twist left, and twist right. 

Agashe et al. \cite{agashe2015global} had decoded hand motions with a different approach. They demonstrated that global cortical activity predicts the shape of the hand during grasping. It was an offline study, and they inferred from EEG hand joint angular velocities as well as synergistic trajectories as subjects perform natural reach-to-grasp movements. They also showed real-time closed-loop neuroprosthetic control of grasping by an amputee and the feasibility of decoding brain signals of a variety of hand motions. However, these related studies could not achieve adequate and robust decoding performance on multiple tasks of the natural hand movements due to its complex characteristics of the brain signal data related to the hand and upper limb. Therefore, we tried to solve this challenging limitation with a new approach and perspective. 

The objective of this study is to confirm whether our proposed method that performs muscle activity pattern matching by creating the estimated muscle activity patterns from each electromyographic (EMG) and EEG signals improves the BCI performance of each subject or not. At the same time, we proved the feasibility of classifying various grasping tasks in the right hand from EEG signals with the proposed method in both actual movement and MI paradigm. Using this new approach, we achieved the improvement of classification accuracy. This approach will be used for further BCI applications, such as controlling a robotic hand. The signals from ten participants were acquired, and we only selected the signals related to muscle activity from the segmented data. With the running of sufficient experimental trials and data analysis, we could construct a robust decoding model based on our proposed method.

\begin{figure}[t!]
\begin{center}
\includegraphics[width = \columnwidth]{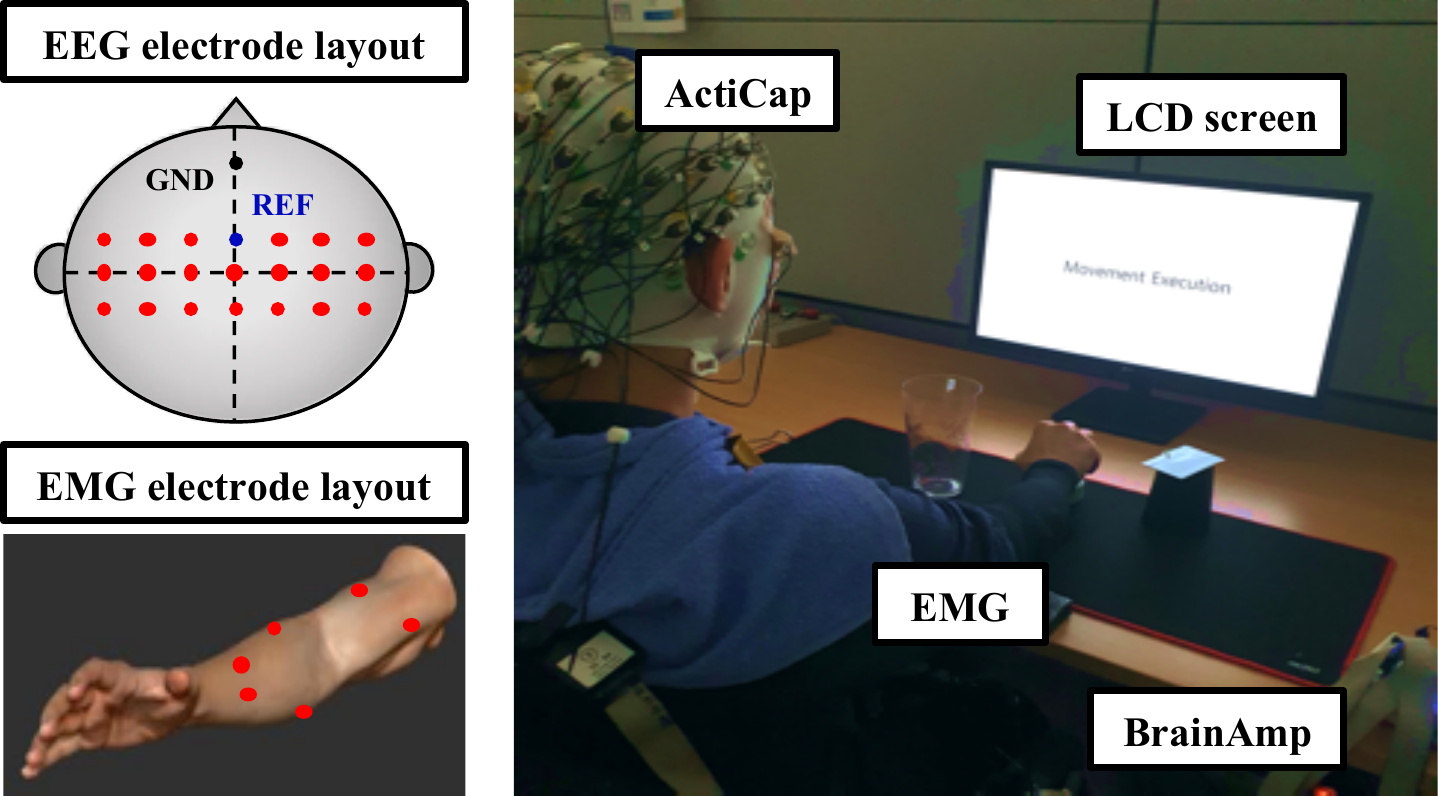}
\end{center}
\caption{Experimental environment and location of EEG and EMG electrodes}
\label{fig1}
\end{figure}

\begin{table}[t]
\centering
\caption{Selected EMG channels and the targeted muscles}
\tiny
\label{table1}
\renewcommand{\arraystretch}{1.2}
\resizebox{\columnwidth}{!}{%
\begin{tabular}{cccc}
\hline
\textbf{Ch.} & \textbf{Target muscle} & \textbf{Related Muscle activity}             \\ \hline
\textbf{1}       & Extensor carpi ulnaris & Wrist extension and abduction        \\
\textbf{2}       & Extensor digitorum     & Finger extension and abduction       \\
\textbf{3}       & Flexor carpi radialis  & Wrist and hand flexion               \\
\textbf{4}       & Flexor carpi ulnaris   & Palm and finger flexion              \\
\textbf{5}       & Biceps brachii         & Forearm lifting                      \\
\textbf{6}       & Triceps brachii        & Forearm extension and retraction     \\ \hline
\end{tabular}%
}
\end{table}

\section{Materials and Methods}

\subsection{Participants}

Ten healthy subjects with no history of neurological disease were recruited for the experiment (S1–S10; ages 24–33; six men, four women; all right-handed). This study was reviewed and approved by the International Review Board, at Korea University [1040548-KU-IRB-17-172-A-2], and written informed consent was obtained from all participants before the experiments. 

\subsection{Experimental Setup}
During a session of the experimental protocol, the subjects sat in front of a 24-inch LCD monitor screen, in a comfortable chair. The screen was installed on the table to make sure that the subjects could see the objects and visual cue. Fig. 1 indicates the experimental setup and the environment during the entire session. The subjects were asked to perform or imagine a specific grasping action following each auditory and visual cue. During the experiment, the subjects were asked to perform three different grasp actions or imagery, which are illustrated in Fig. 2 (a). The location of the object setup was randomly changed to reduce the effect of artifacts.   

\begin{figure}[t!]
\begin{center}
\includegraphics[width=\columnwidth]{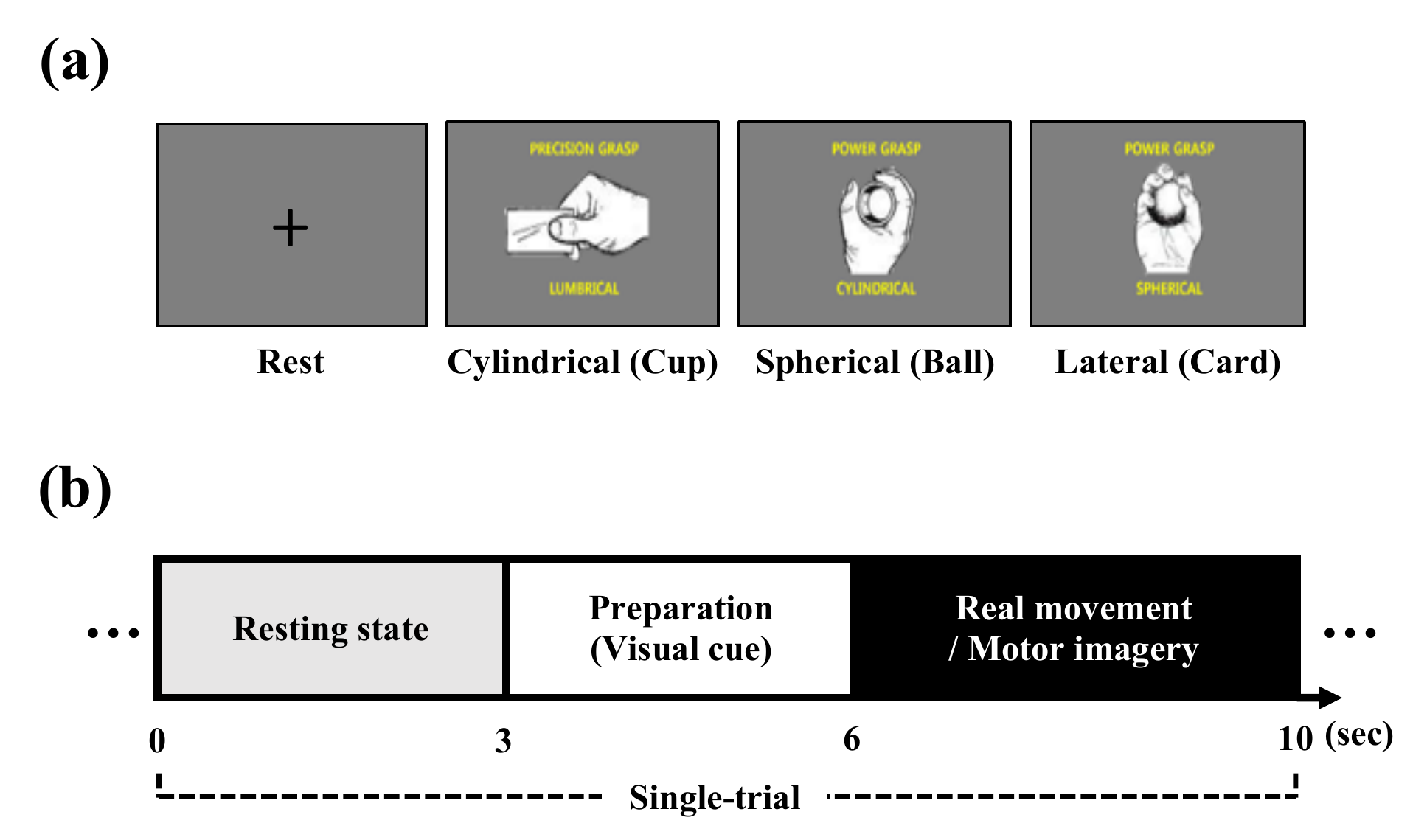}
\end{center}
\caption{Experimental protocol for data acquisition and visual cues}
\label{fig2}
\end{figure}

\begin{figure*}[t!]
\centering
\includegraphics[width=\linewidth]{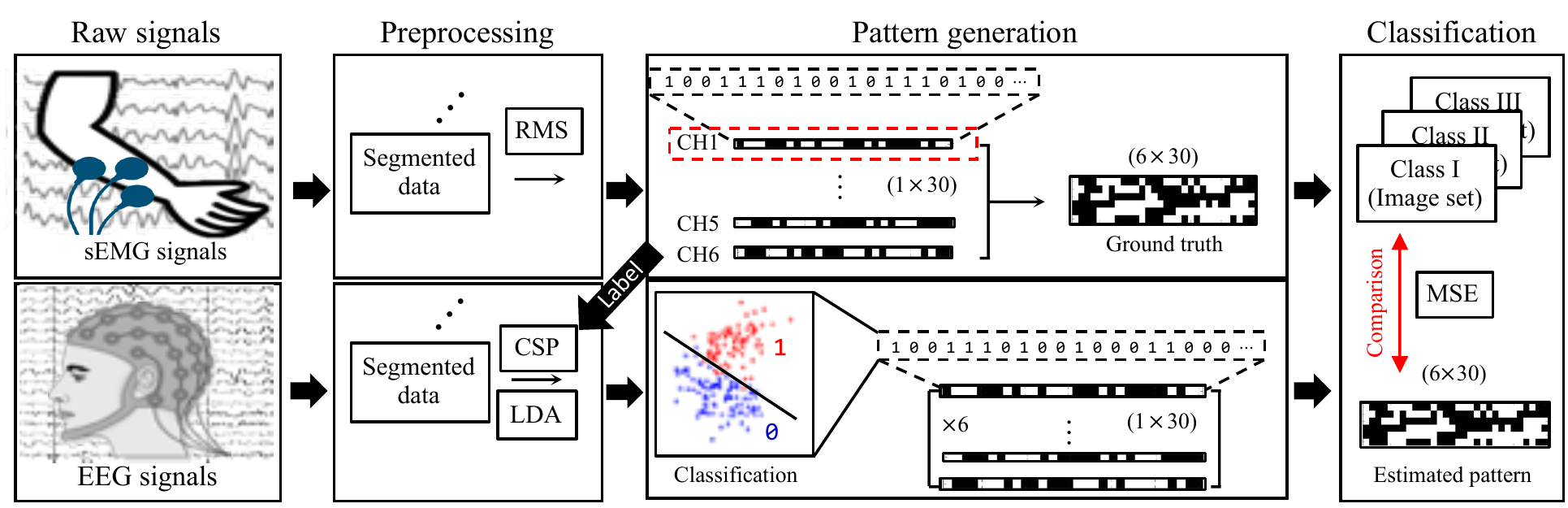}
\caption{Proposed method for classifying grasp actions with muscle activity pattern matching and comparison}
\label{fig3}
\end{figure*}

\begin{figure}[t!]
\begin{center}
\includegraphics[width=\columnwidth]{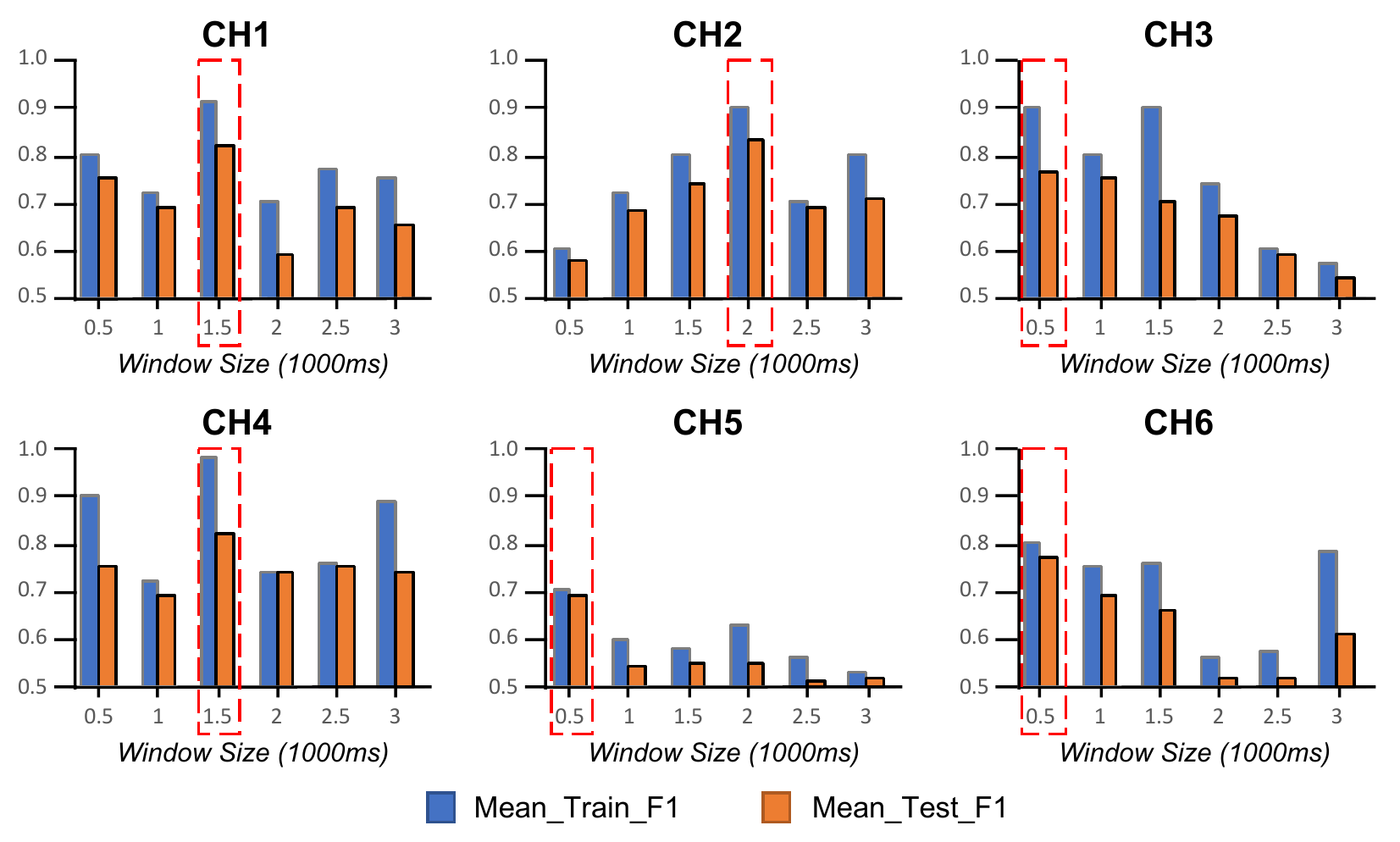}
\end{center}
\caption{Average classification accuracy on the each EMG activation}
\label{fig4}
\end{figure}

\subsection{Data Acquisition}
EEG data were collected at 2,500 Hz using 20 Ag/AgCl electrodes (FC1--6, C1--6, Cz, CP1--6, and CPz) in 10/20 international system via BrainAmp (BrainProduct GmbH)\cite{jeong2018decoding, robinson2013eeg, ang2012filter}. At the same time, a 60 Hz notch filter was used to remove power frequency interference. The FCz and FPz were used as reference and ground electrodes, respectively. All impedances were maintained below 10 \emph{k}$\Omega$. The 20 channels were located only on the motor cortex to make sure that the recorded EEG signals are highly related to the motor-related potentials, which are from the actual movement and MI, as shown in Fig. 1. 

EMG signals were recorded using 7 Ag/AgCl electrodes from a digital amplifier, which is the same equipment to record EEG signals. We acquired the EMG signals with the EEG signals, simultaneously. The EMG data were recorded from six related muscles of right arm movement, as shown in Table I. The last electrode was placed nearby on the right arm elbow, which is a non-muscle movement area, for a reference signal \cite{trigili2019detection}.

\subsection{Data Analysis}
We followed the conventional process of filtering the EEG and EMG signals \cite{cho2018classification, jeong2017single, jeong2018decoding}. We used [4--40] Hz frequency band and separated them into eleven sub-bands for further analysis. The size of the separated bands is 4 Hz size with a step size of 2 Hz \cite{zhang2018temporally, park2017filter}. We extracted spatial patterns using common spatial patterns (CSP) from the filtered EEG \cite{ramoser2000optimal, muller1999designing}. The various window size of 500--2,000 ms was applied to make data segments from the raw EMG signals. The same sliding window was applied to the raw EEG signals as well. After creating the data segments, we adapted the function to calculate the root mean square (RMS) value on the segmented EMG data \cite{li2017motion, trigili2019detection}. Before the RMS step, we already calculated a threshold from the averaged RMS value in a single trial (0--4 sec). When the RMS value of the EMG signal occurs at the specific time point over the pre-defined threshold, we classify this data segment to 0 or 1, a binary classification. We used the sliding windows and made 30 segments from a single trial which is 4,000 ms long. The 1$\times$30 size image shows the result of decoding EMG signals, as shown in Fig 3. This process was performed once by each EMG channel so we could create six images from the six channels. The 6$\times$30-size image shows a muscle activity pattern. We have built the group of pattern for each grasp action class by repeating this process 50 times because each subject performed 50 trials for each class.

In the case of EEG signal decoding, we applied the same moving window from the EMG decoding process. We used this data to train a binary classifier. The label is generated from the EMG decoding and it is used as a training label and ground truth for scoring the performance of the classifier. The CSP was applied to extract spatial features. The features were used as input data to train the linear discriminant analysis (LDA). Throughout the process, we have noted that the estimated pattern from the EEG corresponds to the pattern from the EMG signals. After that, we compared the similarity with the group of patterns, the ground truth. We calculated the mean squared error (MSE) for every 150 pattern images. One of the three that showed the lowest averaged error (the highest similarity) compared to the estimated pattern can be defined as the intended grasp action by the subject.   

\section{Results and Discussion}
The proposed method improves the overall classification performance of the BCI system. We compared our method to the two other competitive models, the model I and model II, as shown in Tables II and III. Model I contains CSP and LDA \cite{ramoser2000optimal, yao2018multi}. Model II contains filter bank regularized CSP (FBRCSP), which usually shows the highest performance in other BCI studies \cite{park2017filter, suk2012novel}. The proposed method that uses muscle activity patterns to classify natural grasp actions showed 24.01\% increased classification accuracy than the model I and 21.59\% improvement compared to the model II.

Table III describes the results in the motor imagery paradigm. The proposed method showed 8.60\% and 5.66\% improvement in the motor imagery paradigm.  It is much less than the result in actual movement, but we point out the proposed method increases the classification performance dramatically on specific subjects, such as S3, 4, and 8. 

In motor imagery, the proposed method showed unstable performance for improving the classification accuracy. We assume that the problem is because of the binary classifier which is used to create the estimated pattern from EEG signals. Unlike the case of the actual movement paradigm, we could not obtain the corresponding EMG signals while the subjects performed motor imagery. Therefore, we recalled the trained classifier from the actual movement decoding process and then applied it to the EEG data of the motor imagery \cite{lee2015subject, jeong2017single}. 

As a result, the similarity of the final estimated pattern as compared to the muscle activity patterns is much lower than the pattern we could create in the actual movement paradigm due to the limitation of estimating without the corresponding EMG signals. Nevertheless, the proposed method also showed an improvement in performance, although the limitation in the motor imagery paradigm.

\begin{table}[t]
\centering
\caption{Classification result comparison in actual movement paradigm}
\tiny
\label{table2}
\renewcommand{\arraystretch}{0.95}
\resizebox{\columnwidth}{!}{%
\begin{tabular}{cccc}
\hline
\multirow{2}{*}{\textbf{Subject}} & \multicolumn{3}{c}{\textbf{Accuracy (\%)}}      \\  
                                  & \textbf{Proposed} & \textbf{Model I} & \textbf{Model II} \\ \hline
\textbf{S1}                       & 69.32            & 39.54            & 41.32             \\
\textbf{S2}                       & 71.98            & 33.04            & 40.12             \\
\textbf{S3}                       & 70.22            & 36.23            & 38.88             \\
\textbf{S4}                       & 67.13            & 42.32            & 41.32             \\
\textbf{S5}                       & 59.10            & 39.31            & 44.08             \\
\textbf{S6}                       & 68.23            & 35.24            & 39.20             \\
\textbf{S7}                       & 47.43            & 48.23            & 49.12             \\
\textbf{S8}                       & 67.02            & 41.48            & 47.33             \\
\textbf{S9}                       & 58.43            & 45.01            & 39.42             \\
\textbf{S10}                      & 60.01            & 37.99            & 42.15             \\ \hline
\textbf{Mean±Std.}                & 63.89±7.54       & 39.84±4.60       & 42.29±3.52        \\ \hline
\end{tabular}%
}
\end{table}

\begin{table}[t]
\centering
\caption{Classification result comparison in motor imagery paradigm}
\tiny
\label{table3}
\renewcommand{\arraystretch}{0.95}
\resizebox{\columnwidth}{!}{%
\begin{tabular}{cccc}
\hline
\multirow{2}{*}{\textbf{Subject}} & \multicolumn{3}{c}{\textbf{Accuracy (\%)}}      \\
                                  & \textbf{Proposed} & \textbf{Model I} & \textbf{Model II} \\ \hline
\textbf{S1}                       & 33.32             & 34.13            & 41.32             \\
\textbf{S2}                       & 38.42             & 33.50            & 36.44             \\
\textbf{S3}                       & 69.23             & 38.12            & 39.43             \\
\textbf{S4}                       & 62.13             & 40.31            & 32.54             \\
\textbf{S5}                       & 38.03             & 40.01            & 35.23             \\
\textbf{S6}                       & 43.23             & 39.31            & 59.32             \\
\textbf{S7}                       & 35.23             & 43.23            & 43.24             \\
\textbf{S8}                       & 73.48             & 32.48            & 44.24             \\
\textbf{S9}                       & 34.11             & 38.49            & 39.32             \\
\textbf{S10}                      & 42.42             & 44.01            & 41.9              \\ \hline
\textbf{Mean±Std.}                & 46.96±15.29       & 38.36±3.93       & 41.30±7.32        \\ \hline
\end{tabular}%
}
\end{table}

\section{Conclusion and Future Work} 
The proposed method suggested a novel approach to decode EEG signals to classify natural grasp actions. In Fig. 4, each graph shows binary classification accuracy for each muscle activity by the EMG channel. Using this binary classifier, we could get reliable estimated muscle activity patterns from EEG data decoding. Our method has the potential to improve performance by increasing the classification accuracy on the binary classifier to get a more accurate estimated muscle activity pattern or improve the similarity comparison step using an advanced model such as deep learning \cite{tabar2016novel, zhang2019novel}.   

\section*{Acknowledgment}
The authors thank to K.-H. Shim, B.-H. Kwon, B.-H. Lee, D.-Y. Lee and D.-H. Lee for help with the database construction and useful discussions of the experiment. 

\ifCLASSOPTIONcaptionsoff
  \newpage
\fi

\bibliographystyle{IEEEtran}
\bibliography{BCI-HandGrasping}

\end{document}